\documentclass[twocolumn,amsmath,amssymb,superscriptaddress,prb]{revtex4}

\usepackage{graphicx}

\usepackage[breaklinks]{hyperref}
\hypersetup{
    bookmarks=false,
    pdfstartview={FitH},
    colorlinks=true,
    linkcolor=blue,
    citecolor=blue,
    urlcolor=blue
}

\begin{document}

\title{{\em k}-resolved susceptibility function of
2{\em H}-TaSe$_2$ from angle-resolved photoemission}

\author{J.~Laverock}
\author{D.~Newby, Jr.}
\author{E.~Abreu}
\author{R.~Averitt}
\author{K.~E.~Smith}
\affiliation{Department of Physics, Boston University, 590 Commonwealth Avenue,
Boston, Massachussetts 02215, USA}

\author{R.~P.~Singh\footnote{Present address:
Department of Physics, IISER Bhopal, MP-462023, India}}
\author{G.~Balakrishnan}
\affiliation{Department of Physics, University of Warwick, Coventry, CV4 7AL,
United Kingdom}

\author{J.~Adell}
\author{T.~Balasubramanian}
\affiliation{MAX-lab, Lund University, SE-221 00 Lund, Sweden}

\begin{abstract}
The connection between the Fermi surface and charge-density wave (CDW)
order is revisited in 2{\em H}-TaSe$_2$. Using angle-resolved photoemission
spectroscopy, {\em ab initio} band structure calculations, and an
accurate tight-binding model, we develop the {\em empirical} {\bf k}-resolved
susceptibility function, which we use to highlight states that contribute to
the susceptibility for a particular {\bf q}-vector. We show
that although the Fermi surface is involved in the peaks in the susceptibility
associated
with CDW order, it is {\em not} through conventional Fermi surface nesting,
but rather
through finite energy transitions from states located far from the Fermi
level. Comparison
with monolayer TaSe$_2$ illustrates the different mechanisms that are
involved in the absence of bilayer splitting.
\end{abstract}

\maketitle

\section{Introduction}
The question of whether nesting instabilities of the Fermi surface (FS)
can drive charge density wave (CDW) formation in real materials has been
the topic of numerous experimental and theoretical investigations for many
years.\cite{wilson1969,gruner1994,whangbo1991}
In cases of apparently well-nested FSs, subsequent
inspection of the real part of the generalized susceptibility, which is the
relevant quantity in assessing instabilities in the electronic system, and
its imaginary counterpart (which is not) can rule against FS nesting being
the {\em primary} driving force.\cite{johannes2008} In concert with
instabilities in the electronic system, lattice effects (through the softening
of phonon modes associated with the CDW) must also be considered on an equal
footing.\cite{calandra2009}

The analysis of the electronic susceptibility of a material is central in
determining whether an electronic instability that may be due to FS nesting is
capable of driving some associated ordering phenomena. Typically, the ${\bf q}$
landscape of the real and imaginary parts of the susceptibility are compared,
and a peak that survives in both parts is taken as evidence that FS nesting may
play a role in emergent phenomena that occurs at that wavevector.  However,
the susceptibility function represents an integral over the Brillouin zone
(BZ), i.e.~over all ${\bf k}$-states. Consequently, one of the deficiencies
of this approach is that some of the most important information that is
available in the susceptibility is integrated out; That is to say, {\em which}
electrons actually contribute to the instability.  In order to illustrate the
importance of the ${\bf k}$-dependence of the susceptibility, we introduce
it here on a prototypical system, 2{\em H}-TaSe$_2$, and demonstrate, from
an experimental perspective, the additional insight that is available from
this kind of analysis.

Of the many CDW materials, the transition metal dichalchogenides are amongst
the most well-known and well-studied.\cite{wilson1969,rossnagel2011}
Indeed, it is
surprising that after the many experimental
\cite{liu1998,straub1999,valla2000,liu2000,tonjes2001,
rossnagel2005,borisenko2008}
and theoretical \cite{rice1975,barnett2006,johannes2008,taraphder2011,ge2012}
investigations, 2{\em H}-TaSe$_2$
still courts controversy as to whether the FS is responsible for its
CDW. Below $T_0 = 122$~K, an incommensurate CDW transition with a wave vector
${\bf q} = (1-\delta)\frac{2}{3}\;{\Gamma}M$ develops, with $\delta \sim 0.02$,
which experiences a lock-in to a commensurate structure ($\delta = 0$)
below 90~K. \cite{moncton1975} The isoelectronic and isostructural compound
2{\em H}-NbSe$_2$, also hosts a similar incommensurate CDW at $T_0 =
33.5$~K.\cite{moncton1975}
Experimentally, the topology of the FS of TaSe$_2$ and NbSe$_2$ are
quantitatively different,
\cite{straub1999,tonjes2001,rossnagel2005}
which immediately raises difficulties with the conventional FS nesting model.
In particular, state-of-the-art bandstructure results firmly rule out
the FS nesting model,\cite{johannes2008} whereas some recent high-resolution
angle-resolved photoemission (ARPES) measurements contradict the theory,
suggesting a primary role for the FS via its experimental autocorrelation
map. \cite{borisenko2008,inosov2008}

Here, we address this controversy
directly through complementary ARPES measurements and {\em ab initio}
bandstructure calculations. Through careful band and {\bf k}-resolved
calculations
of the experimental susceptibility, at energies near and far away from the
Fermi level
($E_{\rm F}$), we show that
FS nesting is too weak to drive CDW order. Instead, peaks in the susceptibility
that are often associated with the CDW originate through finite energy
transitions from bands nested away from $E_{\rm F}$. We show that this
concept explains both the temperature dependent ARPES spectral
function,\cite{borisenko2008,inosov2008} as well as why the material has courted
controversy for so long. Although FS nesting can be ruled out, the Fermi wave
vector, ${\bf k}_{\rm F}$ does play a role, both directly and (more importantly)
indirectly, in determining the peaks in the susceptibility.
We suggest that similar careful inspection of the ${\bf k}$-resolved
susceptibility function in other materials will be capable of discriminating
between different models of charge, spin or superconducting order.

\section{Electronic structure}
\subsection*{{\em Ab initio} calculations}
The electronic structure has been calculated for $2H$-TaSe$_2$ using
the full-potential linear augmented plane-wave (FLAPW) {\sc Elk}
code within the local density approximation (LDA),\cite{elk} including
spin-orbit coupling self-consistently, and using the experimental structural
parameters.\cite{moncton1977}
Relaxation of the unit cell was not found to significantly affect the band
structure, particularly near $E_{\rm F}$.
The band structure of TaSe$_2$ is shown in Fig.~\ref{f:bandsfs}(a),
and is in close agreement with previous electronic structure
calculations.\cite{johannes2008,ge2012}
Two Ta $d$ bands, split by the double TaSe$_2$ layer, cross $E_{\rm F}$ and
form the FS shown on the left of Fig.~\ref{f:bandsfs}(b).
A slight shift downwards
in $E_{\rm F}$ by $\sim 50$~meV, recovers the more familiar FS that has
previously been suggested by
experiment.\cite{rossnagel2005,borisenko2008,inosov2008}
This shifted FS, shown on
the right side of Fig.~\ref{f:bandsfs}(b),
is topologically similar to experiment,
and consists of $\Gamma$- and $K$-centered hole ``barrel'' sheets, from the
first band, and $M$-centered electron ``dogbone'' sheets from the second.
In the following, we refer to the topology of this more familiar, shifted FS.
Note that the downwards shift in $E_{\rm F}$ leads to a reduction in the
band filling of these sheets to 1.80 electrons (from 2).  As pointed out by
Refs.~\onlinecite{rossnagel2007,johannes2008}, relativistic effects are not
negligible for
the heavy Ta ion.  Through time-reversal symmetry, \cite{wexler1976} the
scalar relativistic bands are degenerate across the entire top face ($ALHA$)
of the BZ.
However, with the inclusion of relativistic effects (in the form of spin-orbit
coupling) this restriction is lifted, and the degeneracy is broken.
This
has important, and non-trivial, effects on the FS, allowing the barrel
and dogbone FS sheets to be fully disconnected everywhere (except along
$AL$) in the zone, which ultimately leads to a much more 2D dogbone FS (which,
in turn, ought to enhance the propensity for nesting).

\subsection*{Tight-binding model}
\begin{figure}[t!]
\begin{center}
\includegraphics[width=1.0\linewidth,clip]{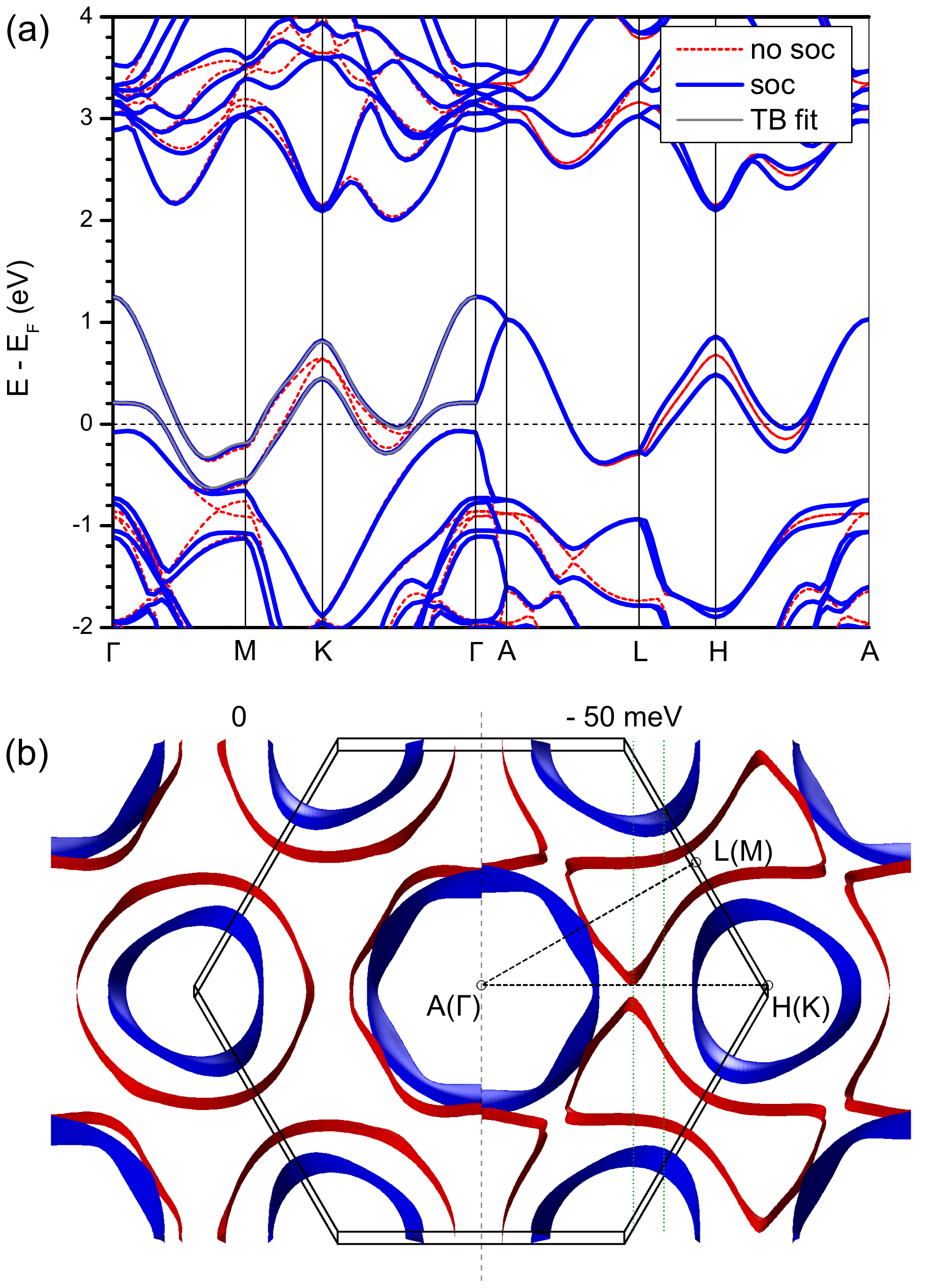}
\end{center}
\vspace*{-0.2in}
\caption{\label{f:bandsfs} (Color online) (a) LDA band structure of 2{\em
H}-TaSe$_2$ including, and neglecting, spin-orbit coupling (soc). The TB$_{\rm
LDA}$ model is shown through $k_z = 0$ in gray.
(b) The FS of 2{\em H}-TaSe$_2$, including soc. The left part shows the raw
LDA FS, and in the right $E_{\rm F}$ has been shifted by $-50$~meV. Symmetry
points in brackets indicate those at $k_z = 0$. The vertical dotted lines
are the slices used in Fig.~\ref{f:chiqk}(c).}
\end{figure}

In order to parameterize the experimental $E({\bf k})$ relation, a simple 2D
tight-binding (TB) model is constructed:
\begin{equation}
E_j({\bf k}) = E_{0,j} + \sum_{\bf R} t_{\lvert {\bf R} \rvert, j}
                \cos ({\bf k}\cdot{\bf R}),
\end{equation}
where ${\bf R}$
are the hexagonal 2D lattice vectors ${\bf a} = \left( \sqrt{3}a/2, \pm a/2
\right)$, $t_{\lvert {\bf R} \rvert}$ are the TB hopping parameters and $j$
is the band index of the two bands that form the FS ($E_0$ is an energy
offset).\cite{barnett2006} In this model, a total of 15 nearest-neighbors
were required to satisfactorily describe a constant $k_z$ slice of the LDA
band structure. Note that a large number of $t_{\lvert {\bf R} \rvert}$ are
used in this work in order to accurately describe both the theoretical and
experimental $E({\bf k})$, and we attach no specific meaning to the individual
parameters.
Before fitting the TB model to the experimental data, we first check its
suitability by assessing how well it is able to describe the theoretical LDA
band structure.
The results of fitting the TB model to the $k_z = 0$ plane
of the LDA band structure (TB$_{\rm LDA}$), shown in Fig.~\ref{f:bandsfs}(a),
agree with the LDA result to within 5~meV~r.m.s.\ in energy.
These TB$_{\rm LDA}$ parameters are only used here to illustrate the capability
of the 15-term TB model in fully capturing the band dispersion of TaSe$_2$,
and its excellent agreement with the {\em ab initio} result demonstrates the
anticipated accuracy of the model in describing the {\em experimental}
dispersion relation. For the remainder of the text, the TB$_{\rm LDA}$
parameters are discarded. Below, we instead carefully fit the experimental
data to the TB model, yielding TB$_{\rm exp}$, which we use for all subsequent
analysis.
Although the model does not explicitly
include spin-orbit coupling, the non-degeneracy of the parameters of the
two bands allow for its effects to be fully captured implicitly.

\subsection*{Angle-resolved photoemission measurements}
Single-crystals of 2{\em H}-TaSe$_2$ were grown by the chemical
vapor transport technique using iodine as the transport
agent.\cite{moncton1977,naito1982etc}
Samples were cleaved in ultra-high vacuum and oriented with
reference to low-energy electron diffraction patterns. Angle-resolved
photoemission measurements were performed at Beamline I4, MAX-lab,
Lund University, Sweden at 100~K with a photon energy of 50~eV and total
instrument resolution of 9~meV. At this temperature, TaSe$_2$ is in the
incommensurate CDW phase, and experiences almost no change in its electronic
structure compared with the normal state.\cite{liu2000,borisenko2008}
The Fermi level was referenced to a gold foil
in electrical contact with the sample.  The experimental dispersion relation
near $E_{\rm F}$ is determined through the 2D curvature of the constant-energy
ARPES intensity map,\cite{zhang2011} $I = I(p_x,p_y)$:
\begin{equation}
C(p_x,p_y) = \frac{(a_0 + I_x^2)I_{yy} - 2I_xI_yI_{xy} + (a_0 + I_y^2)I_{xx}}
{2(a_0+I_x^2+I_y^2)^{\frac{3}{2}}},
\end{equation}
where $I_x = {\partial}I/{\partial}p_x$,
$I_{xx} = {\partial}^2I/{\partial}p_x^2$ and $I_{xy} =
{\partial}^2I/{\partial}p_x{\partial}p_y$ are the partial derivatives of
$I$, and $a_0$ is an arbitrary constant, optimized to maximize the contrast
of $C(p_x,p_y)$. Analysis of the extrema of this function has recently
been shown to accurately locate both band dispersions and FS crossings in
ARPES measurements.\cite{zhang2011} Here, we find it provides significantly
enhanced contrast compared to analysis of the first and/or second derivatives
by themselves, as well as being capable of capturing dispersion parallel to
either direction.

The TB model has been fitted\cite{minuit} to the detected loci to provide
a parameterized description of the experimental $E({\bf k})$ for $E \leq
E_{\rm F}$, which we refer to as TB$_{\rm exp}$.\cite{tbpars}
The energy range of the
fit is restricted to $-260$~meV to $+40$~meV in order to avoid including
the flat portions of the bottom of the bands in the fit; note that these states
are still included in the subsequent analysis.
Experimentally, the bottom
of the lower Ta $d$ band is found to be $-340$~meV, and so most of the band
dispersion is included. In addition to the TB amplitudes, $t_{\lvert {\bf R}
\rvert, j}$, and offsets, $E_{0, j}$, four other adjustable parameters are
varied in the fit, including the lattice parameter, origin in $p_x$ and $p_y$
(projected $\Gamma$-point), and azimuthal alignment, $\theta$.

\begin{figure}[t!]
\begin{center}
\includegraphics[width=0.9\linewidth,clip]{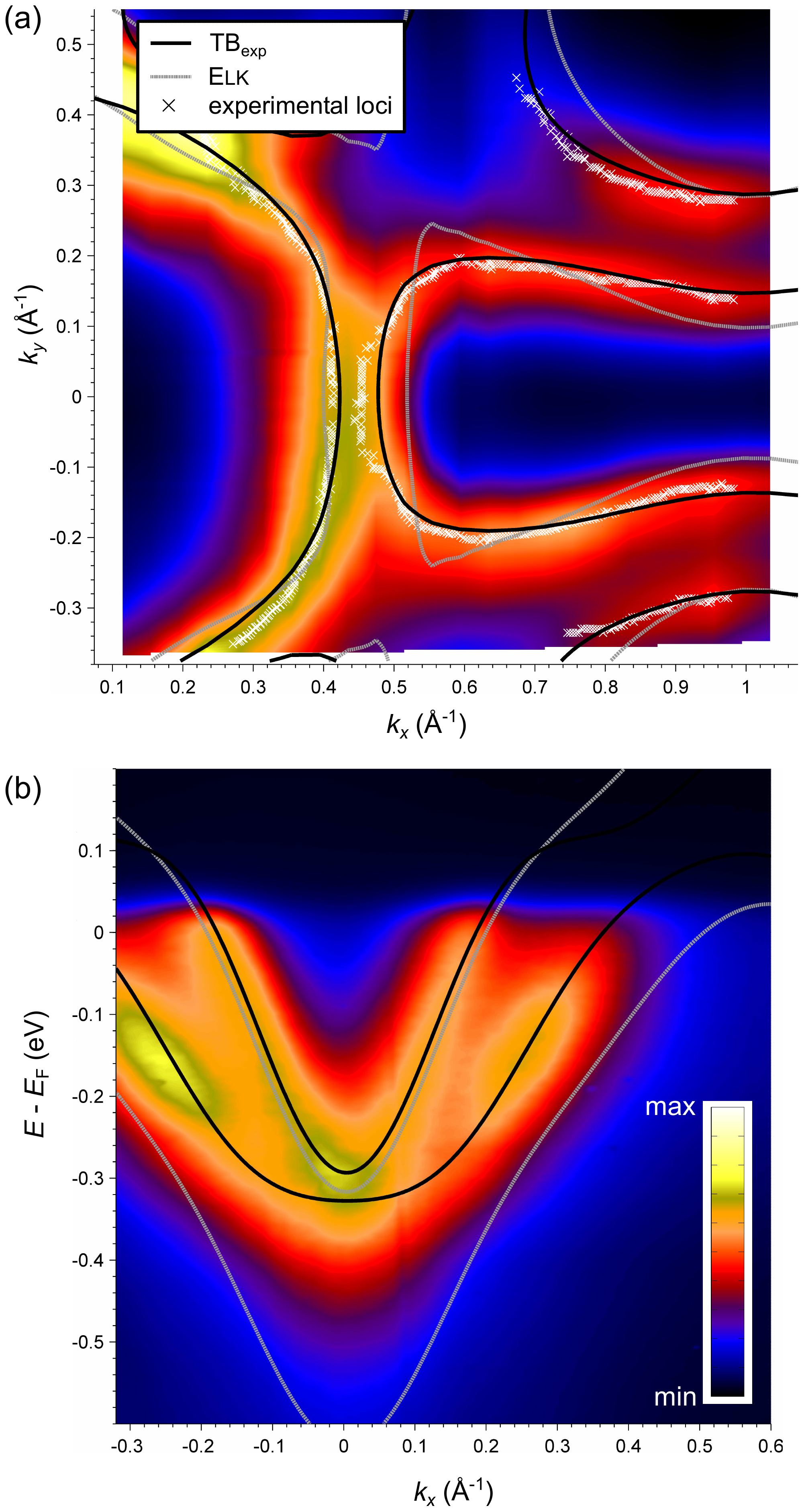}
\end{center}
\vspace*{-0.2in}
\caption{\label{f:arpes} (Color online) (a) ARPES intensity map at $E_{\rm F}$
compared with the FS of the shifted (by $-50$~meV) LDA calculation (light
dashed lines) and of the TB fit to the data (dark solid lines). The detected
band loci are also shown as white crosses. (b) Energy-momentum cut through
$k_x = 0.754$~{\AA}$^{-1}$.}
\end{figure}

The results of the fitted TB$_{\rm exp}$ model are shown in Fig.~\ref{f:arpes}
alongside the ARPES spectra and the shifted LDA result (recall that the raw
calculation yields a topologically different FS). The fit is in excellent
agreement with the data, in both constant energy slices [shown near $E_{\rm
F}$ in Fig.~\ref{f:arpes}(a)] and constant momentum slices [an example is shown
in Fig.~\ref{f:arpes}(b)].  The occupied area of the TB model is 1.92, which is
in closer agreement with the nominal electron count of 2 than the shifted LDA
calculation. This quantity is based on a 2D cut through the 3D band structure,
and is therefore not restricted to obey the Luttinger electron count, but
nevertheless, it is satisfyingly close.  The FS of the TB$_{\rm exp}$ model is
close to previous ARPES
measurements,\cite{borisenko2008,inosov2008,rossnagel2005}
although the $\Gamma$ and $K$ barrels of our FS are slightly smaller and
larger respectively than Refs.~\onlinecite{borisenko2008,inosov2008}. Since this
discrepancy cannot be reproduced by shifts in $E_{\rm F}$, (these sheets are
of the same band), it may reflect a slightly different $k_{\perp}$ associated
with the two different measurements.  Nevertheless, the following analysis
of the data is not affected by changes in ${\bf k}_{\rm F}$ of these sheets,
lending more weight to the argument that FS nesting is weak in TaSe$_2$.

\section{Noninteracting susceptibility}
\subsection*{{\em Ab initio} susceptibility}
\begin{figure}[t!]
\begin{center}
\includegraphics[width=0.9\linewidth,clip]{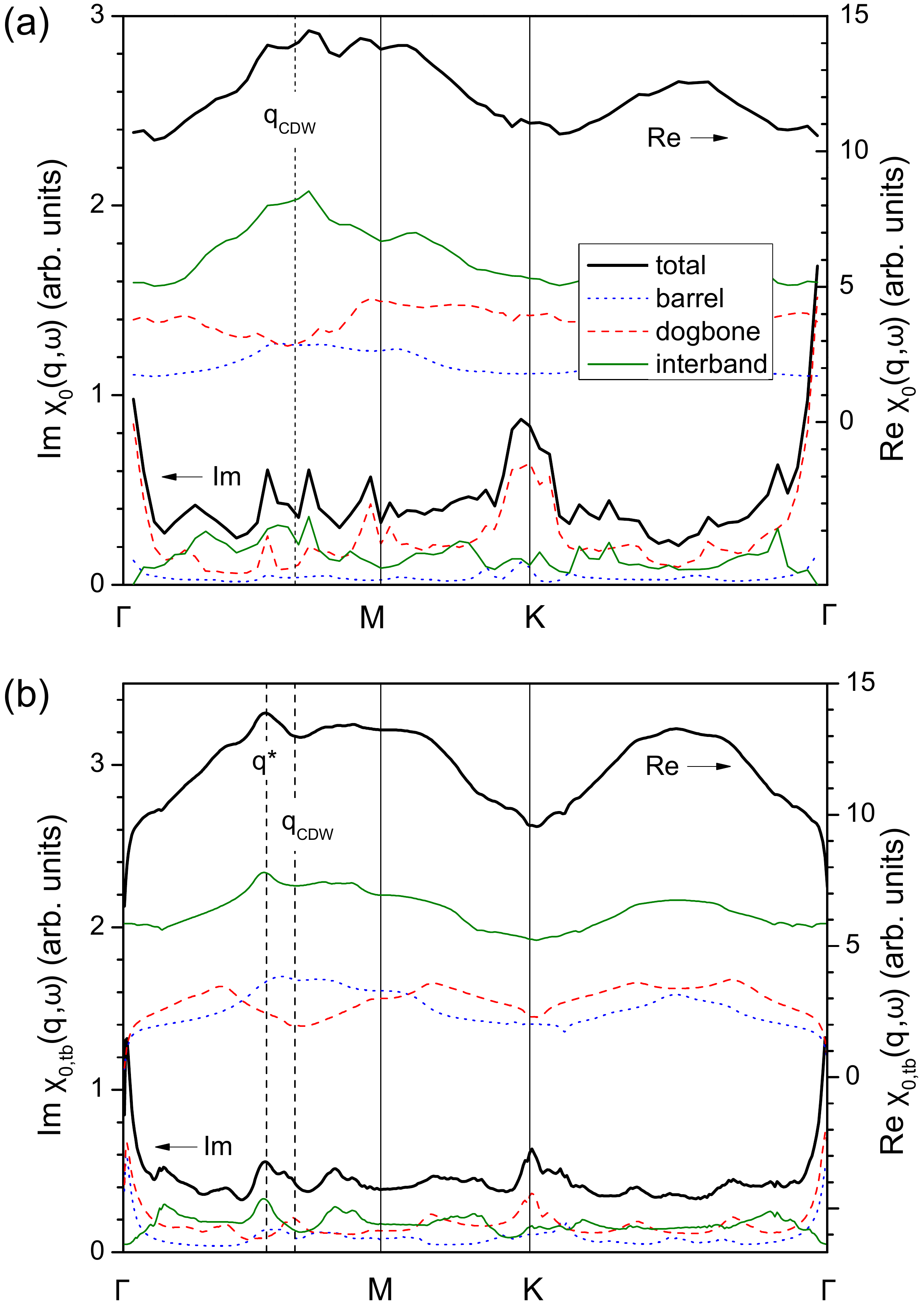}
\end{center}
\vspace*{-0.2in}
\caption{\label{f:chiq} (Color online) Real and imaginary parts of the
noninteracting susceptibility, $\chi_0({\bf q}, \omega)$, for (a) the 3D {\em ab
initio} {\sc Elk} band structure, and (b) the 2D TB$_{\rm exp}$ model bands. The
commensurate CDW wave vector, ${\bf q}_{\rm CDW} = 2/3\;{\Gamma}M$ is indicated
by the dashed
line, and ${\bf q}^* = 0.56\;{\Gamma}M$ indicates the maximum in the TB$_{\rm
exp}$ susceptibility. Note that the real axis is vertically offset for
clarity.}
\end{figure}

The role of nesting in the LDA has
been theoretically investigated via calculations of the noninteracting
susceptibility,
\begin{equation}
\label{e:chiq}
\chi_0({\bf q},\omega) = \sum_{\bf k} \frac{f(\epsilon_{\bf k}) -
f(\epsilon_{{\bf k}+{\bf q}})}{\epsilon_{\bf k} -
\epsilon_{{\bf k}+{\bf q}} - \omega - {\rm i} \delta},
\end{equation}
for wave vector ${\bf q}$ and
frequency $\omega \rightarrow 0$,
in which $f(\epsilon_{\bf k})$ is the Fermi occupancy of
state
$\epsilon_{\bf k}$.\cite{johannes2008,laverock2009} The imaginary part
($\textrm{Im}\,\chi_0$), which gathers transitions in a narrow window
of energies near the FS and can be directly associated with FS nesting,
is shown for TaSe$_2$ in Fig.~\ref{f:chiq}(a) and exhibits some
weak peaks close to, but offset from, ${\bf q}_{\rm CDW}$. The most overwhelming
feature is not at this wavevector, however, but at $q = K$, in which dogbone
nesting dominates.  $\textrm{Im}\,\chi_0$, whilst indicating FS nesting,
is not responsible for CDW order, which instead depends on the real part,
$\textrm{Re}\,\chi_0$.  $\textrm{Re}\,\chi_0$ involves transitions over a
bandwidth-size window of energies, and for TaSe$_2$ is dominated by
interband transitions between the two Ta $d$ bands. The intensity at $K$ is
completely suppressed, and instead $\textrm{Re}\,\chi_0$ peaks at ${\bf q}_{\rm
CDW}$, reflecting the electronic instability that eventually develops
into the CDW. These results, and their interpretation, are very similar
to previous LDA calculations of $\chi_0({\bf q},\omega)$ of TaSe$_2$
and NbSe$_2$.\cite{johannes2006,johannes2008} As we will show below,
from both an experimental {\em and} theoretical perspective, this peak in
$\textrm{Re}\,\chi_0$ has little to do with conventional FS nesting, and
is rather associated with `nesting' between the two bands over energies far from
$E_{\rm F}$.

\begin{figure*}[t!]
\begin{center}
\includegraphics[width=1.0\linewidth,clip]{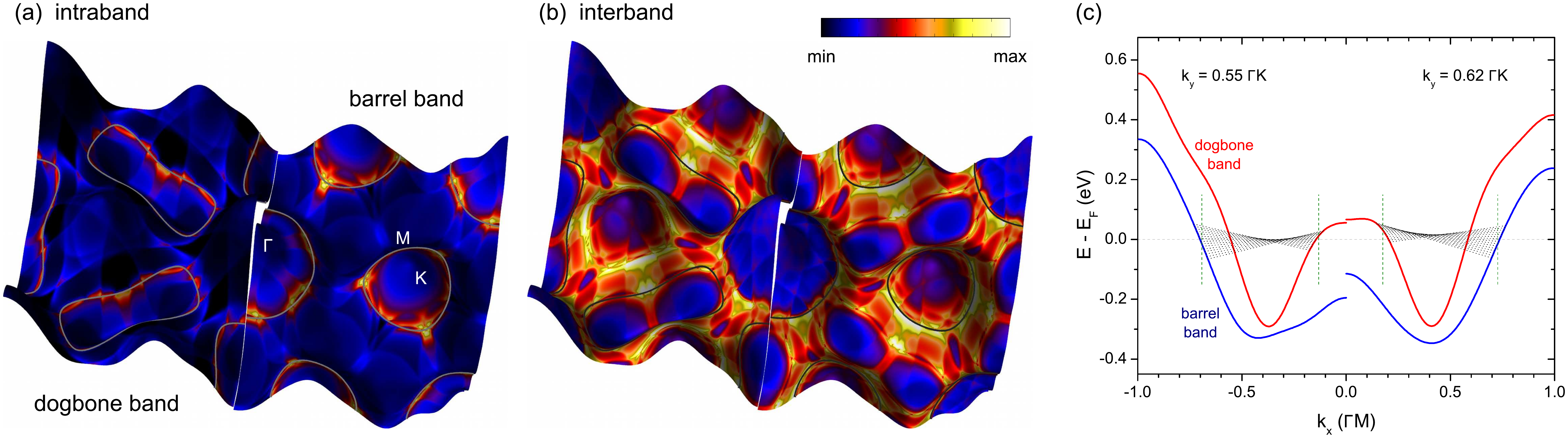}
\end{center}
\vspace*{-0.2in}
\caption{\label{f:chiqk} (Color online) Real part of the ${\bf k}$-resolved
susceptibility function of the TB$_{\rm exp}$ model for ${\bf q} = {\bf q}^*$,
showing (a) intraband 
(dogbone $\rightarrow$ dogbone and barrel $\rightarrow$ barrel) transitions,
and (b) interband (dogbone $\leftrightarrow$ barrel) transitions. The height of
the surface is the TB$_{\rm exp}$ band energy, whereas the color [the same color
scale is used in both (a) and (b)] denotes the magnitude of the
${\bf k}$-resolved susceptibility. The FS of each band is shown
by the gray lines. (c) Slice of the TB$_{\rm exp}$ bands through the
vertical lines of Fig.~\ref{f:bandsfs}(b). The dotted lines all have the same
length of ${\bf q}^*$. Vertical dashed lines indicate the maxima of the ${\bf
k}$-resolved susceptibility for this slice.}
\end{figure*}

\subsection*{Tight-binding susceptibility}
In Fig.~\ref{f:chiq}(b), the noninteracting susceptibility, $\chi_{0, {\rm
tb}}({\bf q}, \omega)$, of the experimental tight-binding model, TB$_{\rm
exp}$, is shown along the same path as the {\em ab initio} result.
This susceptibility, calculated from the experimental band structure, represents
an accurate reflection of the {\em experimental} susceptibility function.
Here,
a temperature of 8~meV is used to fill the states (comparable with the
experiment), although in practice this has little influence on the overall
structure of the susceptibility.  This result, which is based on a 2D
slice of the electronic structure, is of course cruder than the full 3D
calculation shown in Fig.~\ref{f:chiq}(a); nevertheless, the two results
are very similar to one another. In $\textrm{Re}\,\chi_{0, {\rm tb}}$, the
function exhibits a peak near ${\bf q}_{\rm CDW}$, which is predominantly
due to interband transitions. However, the wave vector of this feature is
somewhat offset from ${\bf q}_{\rm CDW}$, rather developing at ${\bf q}^*
= 0.56\;{\Gamma}M$.  Correspondingly, although
$\textrm{Im}\,\chi_{0, {\rm tb}}$
exhibits a weak peak at ${\bf q}^*$ it is neither very intense nor
significantly stronger
than other local peaks elsewhere in the BZ, for example at ${\bf q} = K$,
despite the reduced dimensionality of this 2D model.
In fact, this suppression
of the susceptibility peak is observed in other ARPES
models,\cite{inosov2008,rossnagel2005}
which consistently suggest a slightly lower
${\bf q}$ ($\sim 0.6\;{\Gamma}M$) than the CDW wave vector. This
suggests that ultimately, electron-phonon coupling likely decides which wave
vector is chosen for the ordering.\cite{calandra2009,weber2011} In
all models investigated here, the susceptibility peak is relatively broad and
is certainly compatible with the CDW wave vector.

\subsection*{k-resolved susceptibility}
Unlike typical calculations of the
electronic susceptibility, we now explicitly resolve the ${\bf k}$ dependence
of the susceptibility function,
enabling us to directly assess {\em which} states contribute to $\chi_{0, {\rm
tb}}({\bf q}, \omega)$:
\begin{equation}
\chi_{0, {\rm tb}}({\bf q},{\bf k}) =
\frac{f(\epsilon_{\bf k}) - f(\epsilon_{{\bf k}+{\bf q}})}{\epsilon_{\bf k} -
\epsilon_{{\bf k}+{\bf q}} - \omega - {\rm i} \delta}.
\end{equation}
Here, the integral over the BZ has been dropped with respect to
Eq.~\ref{e:chiq}. For a given value of ${\bf q}$,
this function separates the contribution of each individual
k-point to the susceptibility, allowing the direct visualization in k-space
of which states are connected by that particular wave vector. For example,
for conventional
FS nesting this function will have high intensity only in a narrow region of
k-space near the FS, and will be weak elsewhere.
Integration of
this function over ${\bf k}$ recovers the usual susceptibility function [i.e.\
that shown in Fig.~\ref{f:chiq}(b)].

In Figs.~\ref{f:chiqk}(a,b), $\textrm{Re}\,\chi_{0,{\rm tb}}({\bf q}^*,{\bf k})$
is shown of the experimental TB model (TB$_{\rm exp}$)
for ${\bf q} = {\bf q}^*$. Here,
the magnitude of $\textrm{Re}\,\chi_{0,{\rm tb}}({\bf q}^*,{\bf k})$ is shown as
a color intensity on top of the energy surface of the TB$_{\rm exp}$ bands.
In this presentation, `hotspots' indicate states that are connected to other
states of different occupancy by the wave vector ${\bf q}^*$, and their
intensity reflects their proximity in energy. For reference, the TB$_{\rm exp}$
FS is also shown in Figs.~\ref{f:chiqk}(a,b) as gray contours.

\begin{figure*}[t!]
\begin{center}
\includegraphics[width=1.0\linewidth,clip]{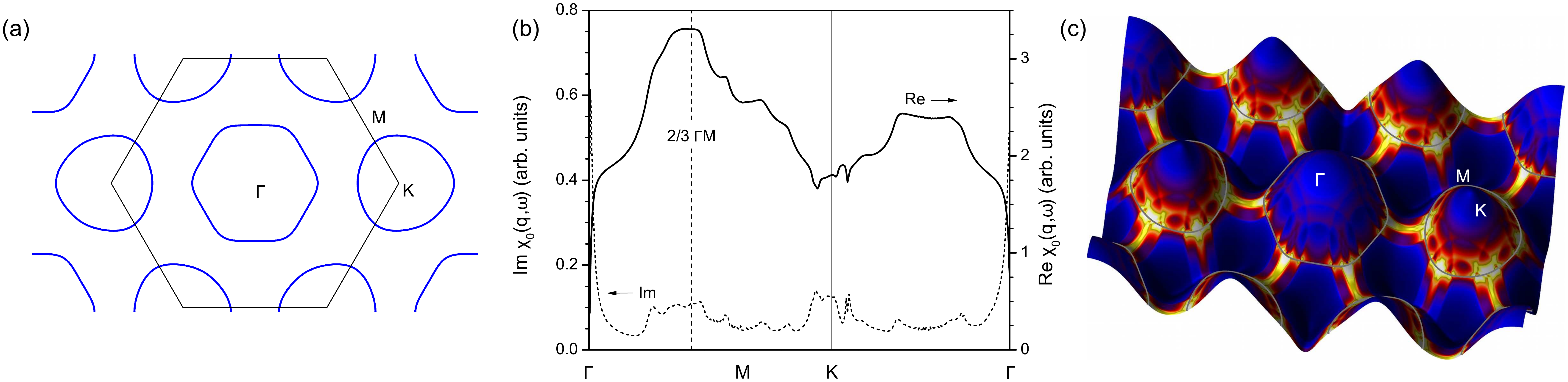}
\end{center}
\vspace*{-0.2in}
\caption{\label{f:monolayer} (Color online) Electronic structure and
susceptibility of monolayer 2{\em H}-TaSe$_2$: (a) Fermi surface, (b)
noninteracting susceptibility, $\chi_0({\bf q}, \omega)$, and (c) real part of
the ${\bf k}$-resolved susceptibility for ${\bf q} = \frac{2}{3}\;{\Gamma}M$.}
\end{figure*}

The intraband contributions [Fig.~\ref{f:chiqk}(a)] of both dogbone (left)
and barrel (right) bands are weak, and
only supply intensity near their respective FSs. It is noted that even though
this function has intensity only near the two FSs, the structure
is `smeared' over a relatively large energy range. Overall, states at least
80~meV away from $E_{\rm F}$ contribute significantly to the intraband
$\chi_{0, {\rm tb}}({\bf q}, \omega)$, which is not compatible with the
conventional FS nesting model.
On the other hand, the interband transitions [Fig.~\ref{f:chiqk}(b)]
show strong intensity over the entire ${\bf k}$ range of the
bands between the two
FSs, irrespective of their energies (which differ by as much as 300~meV in this
part of k-space). This part of the BZ is precisely that
in which the two bands have different occupancies, and therefore in which
transitions are available (through the numerator of Eq.~\ref{e:chiq}).
Similar results are obtained near ${\bf q}^*$ (including at ${\bf q}_{\rm
CDW}$) from the {\em ab initio} unshifted LDA results, despite the different
FS topology, as well as from other TB parameterizations of the energy
bands.\cite{inosov2008} The involvement of such a large region of k-space
in contributing to the susceptibility function, at its peak in ${\bf q}$,
is compelling and direct {\em experimental}
evidence against conventional FS nesting.

\section{Discussion}
Despite our conclusion that FS nesting is not relevant in deciding
the peak in the susceptibility of 2{\em H}-TaSe$_2$, it is evident from
Fig.~\ref{f:chiqk}(b) that there is a reasonable contribution from
interband transitions near the FS,
and it is prudent to ask why this is, given that both
$\textrm{Im}\,\chi_{0, {\rm tb}}$ and the {\em ab initio} $\textrm{Im}\,\chi_0$
clearly rule FS nesting out.
In Fig.~\ref{f:chiqk}(c), two slices of the TB$_{\rm exp}$ energy
bands through $k_y = 0.55\;{\Gamma}K$ and $k_y = 0.62\;{\Gamma}K$ are shown,
corresponding to a vertical slice in Fig.~\ref{f:bandsfs}(b) through both the
$K$ barrels and dogbones. The dotted lines in Fig.~\ref{f:chiqk}(c) all have
the same length in ${\bf k}$-space, {\em viz.}\ ${\bf q}^*$, and connect
unoccupied barrel band states to occupied dogbone band states and {\em vice
versa}. These transitions give rise to the hotspots in Fig.~\ref{f:chiqk}(b)
between the $K$ barrel FS and dogbone FS as well as at the saddle point
along ${\Gamma}K$. The most intense features in $\textrm{Re}\,\chi_{0,
{\rm tb}}$
are shown by vertical dotted lines, and lie in close proximity to
the indicated transitions. Whilst a transition {\em at} the FS is present,
particularly for $k_y = 0.55\;{\Gamma}K$, there is a large number of
finite energy transitions at the same wave vector. The similar magnitude,
but opposite, Fermi velocities of the two bands ensure that this is true
over a large energy range. For $k_y = 0.62\;{\Gamma}K$, the ${\bf q}^*$
vector does not connect to pieces of FS, and instead the FS of each band is
connected to a finite energy {\em away} from the FS of the other band. This
explains why intensity at the FS is visible in Fig.~\ref{f:chiqk}(b), but
very weak in $\textrm{Im}\,\chi_{0, {\rm tb}}$. Although some transitions are
available at the FS, there are many more at finite energy which overwhelm
the low-energy transitions. This concept has similarities, although is more
general, to the idea of ``hidden
nesting'',\cite{whangbo1991} and has been used to explain finite energy
transitions in 1D materials in which FS nesting is `hidden' by band
hybridization effects. We note that
although this explanation was mentioned by Ref.~\onlinecite{johannes2008},
who also categorically ruled out FS nesting, reports of FS nesting-driven
CDW order in the dichalcogenides still pervade the literature.

This explanation of the susceptibility also provides a natural explanation
for why the FS has been implicated in previous studies. To illustrate this,
we consider two bands that have equal and opposite velocities near $E_{\rm F}$,
similar to the case for TaSe$_2$ in Fig.~\ref{f:chiqk}(c),
and which can be idealized in
one dimension as two linear bands of slope $\pm a$ and with Fermi crossings
separated by $k_2 - k_1$. As demonstrated by Johannes {\em et al.}, $\chi({\bf
q})$ can be expanded via $\chi({\bf q}) = \int_{-\infty}^{E_{\rm F}} {\rm d}
x \int^{\infty}_{E_{\rm F}} {\rm d} y\;F(x,y) / (x-y)$, where $F(x,y)=\int
\delta(\epsilon_{\bf k}-x) \delta(\epsilon_{{\bf k}+{\bf q}}-y)\;{\rm d}{\bf
k}$.\cite{johannes2006} The variables $x$ and $y$ relate to states below and
above $E_{\rm F}$. In the idealized 1D model, contributions to $\chi(q)$
are satisfied for $q = (x+y)/a + (k_2-k_1)$, and are weighted by
the energy separation, $x-y$. The integrals over $x$ and $y$, however,
are symmetric about $E_{\rm F}$, and this function must peak at $q =
(k_2-k_1)$, regardless of whether states near $E_{\rm F}$ contribute or
not. Over the full energy window, corresponding to ${\textrm Re}\,\chi(q)$,
the function peaks at this wave vector, not due to FS transitions but rather
due to finite energy transitions (or deep energy nesting).
More generally, in 2D systems this effect is spread out by dispersion over the
second momentum axis, which serves to relax the above idealized arguments.
Nevertheless, it is {\em
not} coincidental that the FS has the same (albeit weak) nesting vector,
but a consequence of the expansion of the transitions about this energy.

This interpretation is consistent with the temperature dependence of the ARPES
spectral function, which becomes gapped (by $\sim 35$~meV) in the commensurate
CDW phase below 90~K.\cite{liu1998,tonjes2001,borisenko2008,demsar2002} The
gapping of the FS occurs most strongly on the $K$ barrels, which completely
disappear, as well as on the long sections of the $M$ dogbone FS (e.g.~the
dogbone crossings along the $KM$ direction). These are precisely the parts
of the FS that were implicated in Fig.~\ref{f:chiqk}(c) as being involved
in the finite energy deep nesting.

To summarize, we have {\em experimentally} shown that the electronic
instability at the FS is not sufficient to drive CDW order in 2{\em
H}-TaSe$_2$. Instead, mechanisms that do not rely on details of the FS are more
likely candidates for driving CDW order. For example, recent models include
the wave vector dependence
of the electron-phonon coupling,\cite{calandra2009,weber2011}
the condensation of preformed excitons,\cite{taraphder2011} or strong
electron-lattice coupling.\cite{gorkov2012}
The importance of analyzing the ${\bf k}$-dependence of the susceptibility
function, at a suitable peak in ${\bf q}$-space, is clearly reflected in our
ability to confidently identify {\em which} electronic states contribute to the
susceptibility at the ordering wavevector. For example, previous experimental
studies, which were based on either the autocorrelation,\cite{borisenko2008}
or a TB fit,\cite{inosov2008} of the ARPES FS (rather than analyzing a large
energy range), concluded that FS nesting was important in driving the CDW of
TaSe$_2$. In contrast, the analysis of the ${\bf k}$-dependence enables us
to firmly rule this out, despite the similarity of our ${\bf k}$-integrated
function to that of Ref.~\onlinecite{inosov2008}, bringing a much-needed
consensus between ARPES experiment and theory.

\section{Monolayer T\lowercase{a}S\lowercase{e}$_2$}
Finally, we consider the situation in the absence of bilayer splitting through
calculations of monolayer TaSe$_2$. In the monolayer, the interband transitions
that were identified in the previous discussion of bulk TaSe$_2$ are not
available, and instead just a single band contributes to the near-$E_{\rm F}$
electronic structure. Moreover, this system is truly 2D, containing no
out-of-plane dispersion, and is therefore more fragile against instabilities in
its FS.

Theoretically, the monolayer is modelled as a single TaSe$_2$ layer
(with the same crystal parameters as the bulk) separated by a vacuum
layer of $\sim 20$~{\AA}, and the {\em ab initio} electronic structure
is calculated using the FLAPW {\sc Elk} code. No attempt has been
made to relax the structural parameters. The FS of monolayer TaSe$_2$
is shown in Fig.~\ref{f:monolayer}(a), and consists of a rounded
hexagon centered at ${\Gamma}$ and a rounded triangle at $K$, similar
in topology to previous results on monolayer TaSe$_2$\cite{ge2012}
and monolayer NbSe$_2$.\cite{calandra2009} In Fig.~\ref{f:monolayer}(b),
the susceptibility of this band structure is shown (and is in good agreement
with previous calculations).\cite{ge2012}  The real part exhibits a strong peak
centered close to $\frac{2}{3}\;{\Gamma}M$, although, similar to the bulk,
the peak in the imaginary part remains broad and smaller than at ${\bf q}
= K$. In (relaxed) monolayer NbSe$_2$, the peak in the susceptibility was
found to shift to $\frac{1}{2}\;{\Gamma}M$.\cite{calandra2009}

The ${\bf k}$-resolved susceptibility is shown in Fig.~\ref{f:monolayer}(c)
for ${\bf q} = \frac{2}{3}\;{\Gamma}M$.  In the absence of interband
transitions, the peak in the susceptibility of the monolayer is associated
with the saddle points of the band structure, which connect to the vicinity
of the $K$ FS sheet.  This situation is reminiscent, although quantitatively
different, to the saddle point nesting model, which was based on a single
NbSe$_2$ band.\cite{rice1975} Indeed, a more recent ARPES study postulated
that both the saddle band and the $K$ FS may be involved.\cite{tonjes2001}
The peaks in both bulk and monolayer susceptibilities involve states near
the saddle band region and near the $K$ FS barrels, and it is this ${\bf
q}$-vector that is most relevant in determining the susceptibility peak.
However, the ${\bf k}$-dependence of the susceptibility is quite
different in the monolayer, being restricted to narrow strips near the
saddle band region. These results demonstrate the sensitivity of the ${\bf
k}$-resolved susceptibility to changes in the active states at a
particular ${\bf q}$-vector, and illustrate its value in assessing the origin of
instabilities in the electronic subsystem.

\section{Conclusion}
The connection between the FS and the CDW has been revisited in
2{\em H}-TaSe$_2$ through ARPES measurements. After developing an
accurate tight-binding model of the experimental electronic structure,
the experimental susceptibility was calculated, and compared with {\em ab
initio}
calculations.  Through careful analysis of the empirical ${\bf k}$-resolved
electronic susceptibility function, finite energy transitions have been shown
to dominate the susceptibility both at its peak, and at the CDW wave vector.
This approach directly illustrates {\em which} states are involved in features
of the electronic susceptibility. Whilst the conventional FS nesting model
is considered too weak to drive the CDW, the FS is indirectly involved in
determining the peak in the susceptibility, although the final choice of
ordering vector likely depends on the lattice. Finally, comparison with
theoretical calculations of 2D monolayer TaSe$_2$ illustrate the different
electron states that are involved in the absence of bilayer splitting.

\section*{Acknowledgements}
We would like to thank S.\ B.\ Dugdale and A.\ R.\ H.\ Preston for valuable
discussions.
The Boston University program is supported in part by the
Department of Energy under Grant No.\ DE-FG02-98ER45680.
The work in the UK was supported by EPSRC, UK, (EP/I007210/1) and the Boston
University/University of Warwick collaboration fund.
EA acknowledges support from Funda\c{c}\~{a}o para a Ci\^{e}ncia e a
Tecnologia, Portugal, through a
doctoral degree fellowship (SFRH/BD/47847/2008).


\begin{thebibliography}{99}

\bibitem{wilson1969}
J.\ A.\ Wilson and A.\ D.\ Yoffe,
\href{http://dx.doi.org/10.1080/00018736900101307}
{Adv.\ Phys.\ {\bf 18}, 193 (1969)}.

\bibitem{gruner1994}
G.\ Gr\"{u}ner, {\em Density Waves in Solids} (Addison-Wesley, Reading, PA,
1994).

\bibitem{whangbo1991}
M.-H.\ Whangbo, E.\ Canadell, P.\ Foury and J.-P.\ Pouget,
\href{http://dx.doi.org/10.1126/science.252.5002.96}
{Science {\bf 252}, 96 (1991)}.

\bibitem{johannes2008}
M.\ D.\ Johannes and I.\ I.\ Mazin,
\href{http://dx.doi.org/10.1103/PhysRevB.77.165135}
{Phys.\ Rev.\ B {\bf 77}, 165135 (2008)}.

\bibitem{calandra2009}
M.\ Calandra, I.\ I.\ Mazin and F.\ Mauri,
\href{http://dx.doi.org/10.1103/PhysRevB.80.241108}
{Phys.\ Rev.\ B {\bf 80}, 241108(R) (2009)}.

\bibitem{rossnagel2011}
K.\ Rossnagel,
\href{http://dx.doi.org/10.1088/0953-8984/23/21/213001}
{J.\ Phys.: Condens.\ Matter {\bf 23}, 213001 (2011)}.

\bibitem{liu1998}
R.\ Liu, C.\ G.\ Olson, W.\ C.\ Tonjes and R.\ F.\ Frindt,
\href{http://dx.doi.org/10.1103/PhysRevLett.80.5762}
{Phys.\ Rev.\ Lett.\ {\bf 80}, 5762 (1998)}.

\bibitem{valla2000}
T.\ Valla, A.\ V.\ Fedorov, P.\ D.\ Johnson, J.\ Xue, K.\ E.\ Smith and
F.\ J.\ DiSalvo,
\href{http://dx.doi.org/10.1103/PhysRevLett.85.4759}
{Phys.\ Rev.\ Lett.\ {\bf 85}, 4759 (2000)};
T.\ Valla, A.\ V.\ Fedorov, P.\ D.\ Johnson, P.-A.\ Glans, C.\ McGuinness,
K.\ E.\ Smith, E.\ Y.\ Andrei and H.\ Berger,
\href{http://dx.doi.org/10.1103/PhysRevLett.92.086401}
{Phys.\ Rev.\ Lett.\ {\bf 92}, 086401 (2004)}.

\bibitem{liu2000}
R.\ Liu, W.\ C.\ Tonjes, V.\ A.\ Greanya, C.\ G.\ Olson and R.\ F.\ Frindt,
\href{http://dx.doi.org/10.1103/PhysRevB.61.5212}
{Phys.\ Rev.\ B {\bf 61}, 5212 (2000)}.

\bibitem{straub1999}
Th.\ Straub, Th.\ Finteis, R.\ Claessen, P.\ Steiner, S.\ H\"{u}fner, P.\ Blaha,
C.\ S.\ Oglesby and E.\ Bucher,
\href{http://dx.doi.org/10.1103/PhysRevLett.82.4504}
{Phys.\ Rev.\ Lett.\ {\bf 82}, 4504 (1999)}.

\bibitem{tonjes2001}
W.\ C.\ Tonjes, V.\ A.\ Greanya, R.\ Liu, C.\ G.\ Olson and P.\ Molini\'{e},
\href{http://dx.doi.org/10.1103/PhysRevB.63.235101}
{Phys.\ Rev.\ B {\bf 63}, 235101 (2001)}.

\bibitem{rossnagel2005}
K.\ Rossnagel, E.\ Rotenberg, H.\ Koh, N.\ V.\ Smith and L.\ Kipp,
\href{http://dx.doi.org/10.1103/PhysRevB.72.121103}
{Phys.\ Rev.\ B {\bf 72}, 121103(R) (2005)}.

\bibitem{borisenko2008}
S.\ V.\ Borisenko, A.\ A.\ Kordyuk, A.\ N.\ Yaresko, V.\ B.\ Zabolotnyy,
D.\ S.\ Inosov, R.\ Schuster, B.\ B\"{u}chner, R.\ Weber, R.\ Follath,
L.\ Patthey and H.\ Berger,
\href{http://dx.doi.org/10.1103/PhysRevLett.100.196402}
{Phys.\ Rev.\ Lett.\ {\bf 100}, 196402 (2008)}.

\bibitem{rice1975}
T.\ M.\ Rice and G.\ K.\ Scott,
\href{http://dx.doi.org/10.1103/PhysRevLett.35.120}
{Phys.\ Rev.\ Lett.\ {\bf 35}, 120 (1975)}.

\bibitem{barnett2006}
R.\ L.\ Barnett, A.\ Polkovnikov, E.\ Demler, W.-G.\ Yin and W.\ Ku,
\href{http://dx.doi.org/10.1103/PhysRevLett.96.026406}
{Phys.\ Rev.\ Lett.\ {\bf 96}, 026406 (2006)}.

\bibitem{taraphder2011}
A.\ Taraphder, S.\ Koley, N.\ S.\ Vidhyadhiraja and M.\ S.\ Laad,
\href{http://dx.doi.org/10.1103/PhysRevLett.106.236405}
{Phys.\ Rev.\ Lett.\ {\bf 106}, 236405 (2011)}.

\bibitem{ge2012}
Y.\ Ge and A.\ Y.\ Liu,
\href{http://dx.doi.org/10.1103/PhysRevB.86.104101}
{Phys.\ Rev.\ B {\bf 86}, 104101 (2012)}.

\bibitem{moncton1975}
D.\ E.\ Moncton, J.\ D.\ Axe and F.\ J.\ DiSalvo,
\href{http://dx.doi.org/10.1103/PhysRevLett.34.734}
{Phys.\ Rev.\ Lett.\ {\bf 34}, 734 (1975)}.

\bibitem{inosov2008}
D.\ S.\ Inosov, V.\ B.\ Zabolotnyy, D.\ V.\ Evtushinsky, A.\ A.\ Kordyuk,
B.\ B\"{u}chner, R.\ Follath, H.\ Berger and S.\ V.\ Borisenko,
\href{http://dx.doi.org/10.1088/1367-2630/10/12/125027}
{New J.\ Phys.\ {\bf 10}, 125027 (2008)}.

\bibitem{elk}
J.\ K.\ Dewhurst, S.\ Sharma, L.\ Nordst\"{o}m, F.\ Cricchio, F.\ Bultmark and
E.\ K.\ U.\ Gross,
\href{http://elk.sourceforge.net}
{\tt{http://elk.sourceforge.net}} (2012).

\bibitem{moncton1977}
D.\ E.\ Moncton, J.\ D.\ Axe and F.\ J.\ DiSalvo,
\href{http://dx.doi.org/10.1103/PhysRevB.16.801}
{Phys.\ Rev.\ B {\bf 16}, 801 (1977)}.

\bibitem{rossnagel2007}
K.\ Rossnagel and N.\ V.\ Smith,
\href{http://dx.doi.org/10.1103/PhysRevB.76.073102}
{Phys.\ Rev.\ B {\bf 76}, 073102 (2007)}.

\bibitem{wexler1976}
G.\ Wexler and A.\ M.\ Woolley,
\href{http://dx.doi.org/10.1088/0022-3719/9/7/010}
{J.\ Phys.\ C: Solid State Phys.\ {\bf 9}, 1185 (1976)}.

\bibitem{naito1982etc}
M.\ Naito and S.\ Tanaka,
\href{http://dx.doi.org/10.1143/JPSJ.51.219}
{J.\ Phys.\ Soc.\ Jpn.\ {\bf 51}, 219 (1982)};
E.\ Revolinsky, B.\ E.\ Brown, D.\ J.\ Beerntsen and C.\ H.\ Armitage,
\href{http://dx.doi.org/10.1016/0022-5088(65)90058-5}
{J.\ Less-Common Met.\ {\bf 8}, 63 (1965)}.

\bibitem{zhang2011}
P.\ Zhang, P.\ Richard, T.\ Qian, Y.-M.\ Xu and X.\ Dai,
\href{http://dx.doi.org/10.1063/1.3585113}
{Rev.\ Sci.\ Instrum.\ {\bf 82}, 043712 (2011)}.

\bibitem{minuit}
F.~James, CERN Program Library D506, MINUIT--Function Minimization and Error
Analysis, 1994, {\tt http://consult.cern.ch/writeups/minuit},
{\tt http://c-minuit.sourceforge.net}.

\bibitem{tbpars}
Taking $i$ as the index into the ordered set, $\lvert {\bf R} \rvert$, the
$E_{0,j}$ where $j$ is the band index, and $t_{i,j}$ ($1 \leq i \leq 15$),
are (in meV): $E_{0,1}$ = $-79.4$; $t_{i,1}$ = $20.5$, $74.1$, $-12.0$, $-2.9$,
$-3.1$, $-0.1$, $0.3$, $4.2$, $2.5$, $3.1$, $3.3$, $5.1$, $2.8$, $0.6$, $2.3$;
$E_{0,2}$ = $118.1$; $t_{i,2}$ = $35.9$, $91.6$, $-3.6$, $6.5$, $-8.7$, $2.9$,
$5.8$, $-11.9$, $-5.0$, $5.3$, $-3.5$, $-1.0$, $-0.4$, $1.5$, $-0.3$.

\bibitem{laverock2009}
J.\ Laverock, T.\ D.\ Haynes, C.\ Utfeld and S.\ B.\ Dugdale,
\href{http://dx.doi.org/10.1103/PhysRevB.80.125111}
{Phys.\ Rev.\ B {\bf 80}, 125111 (2009)}.

\bibitem{johannes2006}
M.\ D.\ Johannes, I.\ I.\ Mazin and C.\ A.\ Howells,
\href{http://dx.doi.org/10.1103/PhysRevB.73.205102}
{Phys.\ Rev.\ B {\bf 73}, 205102 (2006)}.

\bibitem{weber2011}
F.\ Weber, S.\ Rosenkranz, J.-P.\ Castellan, R.\ Osborn, R.\ Hott, R.\ Heid,
K.-P.\ Bohnen, T.\ Egami, A.\ H.\ Said and D.\ Reznik,
\href{http://dx.doi.org/10.1103/PhysRevLett.107.107403}
{Phys.\ Rev.\ Lett.\ {\bf 107}, 107403 (2011)}.

\bibitem{demsar2002}
J.\ Demsar, L.\ Forr\'{o}, H.\ Berger and D.\ Mihailovic,
\href{http://dx.doi.org/10.1103/PhysRevB.66.041101}
{Phys.\ Rev.\ B {\bf 66}, 041101(R) (2002)}.

\bibitem{gorkov2012}
L.\ P.\ Gor'kov,
\href{http://dx.doi.org/10.1103/PhysRevB.85.165142}
{Phys.\ Rev.\ B {\bf 85}, 165142 (2012)}.

\end{thebibliography}
\end{document}